\documentclass{PoS}
\usepackage{amsmath}

\title{Gauge and Higgs Boson Masses from an Extra Dimension}

\ShortTitle{Gauge and Higgs Boson Masses from an Extra Dimension}

\author{\speaker{Graham Moir}, Peter Dziennik, Francesco Knechtli and Kyoko Yoneyama \\
        Department of Physics, Bergische Universit\"{a}t Wuppertal, Gaussstr. 20, D-42119 Wuppertal, Germany \\
       E-mail: \email{moirg@tcd.ie}, \email{knechtli@uni-wuppertal.de}}
\author{Nikos Irges \\
Department of Physics, National Technical University of Athens Zografou Campus, GR-15780 Athens, Greece \\
\email{irges@mail.ntua.gr}\\
}

\abstract{ We present novel calculations of the mass hierarchy of the $SU(2)$ pure gauge theory on a space-time lattice with an orbifolded fifth dimension. This theory has three parameters; the gauge coupling $\beta$, the anisotropy $\gamma$, which is a measure of the ratio of the lattice spacing in the four dimensions to that in the fifth dimension, and the extent of the extra dimension $N_{5}$. Using a large basis of scalar and vector operators we explore in detail the spectrum along the $\gamma = 1$ line, and for the first time we investigate the spectrum for $\gamma \neq 1$.}

\FullConference{The 32nd International Symposium on Lattice Field Theory\\
                 23-28 June, 2014\\
                 Columbia University New York, NY}

\begin{document}

\section*{Introduction}

The recent discovery of a scalar particle around $125$ GeV at the LHC has all but confirmed the existence of the Higgs mechanism, rendering the Standard Model of particle physics complete \cite{ATLAS:2012gk,CMS:2012gu}. However, the origin of this mechanism is, as of yet, unknown and the arbitrary nature of the required fine tuning of parameters suggests that a more fundamental process is at work. One exciting explanation is that of Gauge-Higgs Unification (GHU), in which space-time is extended to include compactified extra-dimensions and the Higgs field is identified with the extra-dimensional components of the gauge field. When the extra-dimensional space is not simply connected, spontaneous symmetry breaking (SSB) of the vacuum is induced dynamically solving the aforementioned fine tuning problem. 

The GHU model discussed here is that of a five-dimensional $SU(2)$ gauge theory where the extra dimension is compactified on an $S^{1}/\mathbb{Z}_{2}$ orbifold. At the fixed points of the orbifold, the gauge group is broken down to $U(1)$ and the theory exhibits SSB \cite{Irges:2006hg} in accordance with Elitzur's theorem \cite{Elitzur}, via the spontaneous breaking of the co-called \textit{stick symmetry} \cite{Ish,Irges:2013rya}. A significant feature of this theory is that is exhibits SSB without the presence of fermionic degrees of freedom, which are required perturbatively by the Hosotani mechanism \cite{Hosotani:1983vn}. We label our five-potential $A_{M}$, where $M = 0, 1, 2, 3, 5,$ corresponding to the usual four dimensions and an extra fifth dimension. The $A^{1,2}_{5}$ components of the gauge field are identified with a Higgs-like complex scalar and the $A^{3}_{5}$ component is identified with a $Z$-like vector boson. 

Clearly, since we are working in five dimensions, the theory is non-renormalisable. However, since we can scan the phase diagram in a non-perturbative fashion, it is possible to probe the system in the vicinity of a bulk phase transition where a scaling regime with suppressed cut-off effects may exist.

\section*{Lattice Set-Up}

We impose orbifold boundary conditions on the extra dimension following \cite{Irges:2006hg,Irges:2004gy}. One first considers a gauge theory constructed on a five-dimensional periodic Euclidean lattice $T \times L^{3} \times N_{5}$, where $T$ refers to the temporal extent of the lattice, $L$ to the spatial extent and $N_{5}$ to the extent of the extra dimension. We denote the lattice spacing as $a$ and label the coordinates of the lattice points via a set of integers $n \equiv \{n_{M}\}$. In what follows, $\mu$ is used to index the standard four dimensions. The gauge field is defined as the set of gauge links $\{U(n,M) \in SU(2)\}$ that connect neighbouring lattice points. The orbifolding of the extra dimension is then achieved by a combination of a reflection $\mathcal{R}$ and group conjugation $\mathcal{T}_{g}$, hence leading us to the orbifolding condition
\begin{equation}\label{eqn:orbifold_condition}
(1 - \mathcal{RT}_{g}) U(n,M) = 0~,
\end{equation}
which is essentially a $\mathbb{Z}_{2}$ projection of the gauge links. The reflection operation acts on both the lattice points 
\begin{equation} 
\bar{n} \equiv \mathcal{R}n = (n_{\mu}, -n_{5})~,
\end{equation}
and the gauge links
\begin{equation}
\mathcal{R}U(n,\mu) = U(\bar{n}, \mu) ~~~,~~~ \mathcal{R}U(n, 5) = U^{\dagger}(\bar{n} - \hat{5}, 5)~.
\end{equation}
The group conjugation acts solely on the gauge links
\begin{equation}
\mathcal{T}_{g}U(n,M) = g U(n,M) g^{-1}~,
\end{equation}
where $g$ is a constant $SU(2)$ matrix such that $g^{2} \in Z(SU(2))$. We \textit{choose} $g = -i\sigma^{3}$, and in accordance with gauge invariance, the gauge group is broken down to $U(1)$ at the fixed points of the orbifold ($n_{5} = 0, N_{5}$). 
\begin{figure}[t!]
  \begin{center}
   \includegraphics[width=0.7\textwidth]{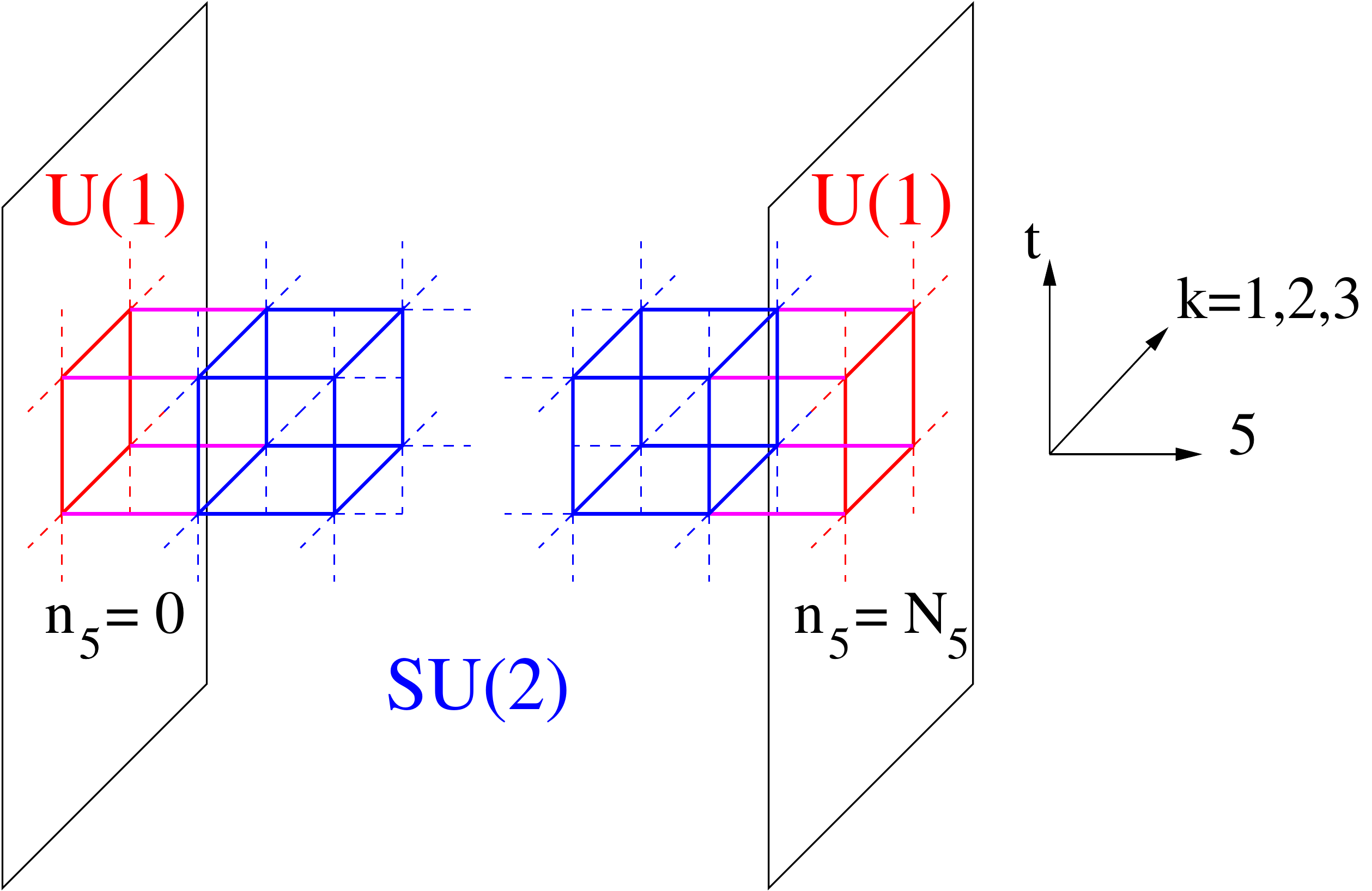}
  \end{center}
  \caption{\footnotesize Sketch of the orbifold lattice and the gauge links: boundary $U(1)$ links are red, hybrid five-dimensional $SU(2)$ links
sticking to the boundaries are magenta and bulk $SU(2)$ links are blue.}
  \label{fig:latorb}
\end{figure}
The theory is now defined in the domain $I = \{ n_{\mu}, 0 \leq n_{5} \leq N_{5} \}$ corresponding to Figure \ref{fig:latorb}. We perform our study using a five-dimensional anisotropic Wilson action
\begin{equation}
S_W^{orb} = \frac{\beta}{2} \sum_{n_{\mu}} {\left[ \frac{1}{\gamma}\sum_{\mu<\nu}{w~\text{tr}{\left\{1 - U_{\mu\nu}(n_{\mu}) \right\}}} + \gamma \sum_\mu{\text{tr}{\left\{ 1 - U_{\mu5}(n_{\mu}) \right\}}} \right]}~,
\end{equation}
where $w = 1/2$ for plaquettes, $U_{\mu \nu}$, living at the fixed points of the orbifold and $w = 1$ otherwise. The anisotropy parameter $\gamma = a_{4}/a_{5}$ in the classical continuum limit, where $a_{4}$ denotes the lattice spacing in the usual four dimensions and $a_{5}$ denotes the lattice spacing along the extra dimension. As depicted in Figure \ref{fig:latorb}, there are three types of gauge links: four-dimensional $U(1)$ links residing at the fixed points of the orbifold, bulk $SU(2)$ links and extra dimensional links which have one end at a fixed point and the other in the bulk; these so-called \textit{hybrid} links gauge transform as $U \rightarrow \Omega^{U(1)} U \Omega^{\dag SU(2)}$ at the left boundary ($n_{5} = 0$) and $U \rightarrow \Omega^{SU(2)} U \Omega^{\dag U(1)}$ at the right boundary ($n_{5} = N_{5}$).

\section*{Mass Extraction}

We obtain spectral information from two-point correlation functions
\begin{equation}\label{eqn:correlation_function}
C_{ij}(t) = \langle \mathcal{O}_{i}(t) \mathcal{O}^{\dagger}_{j}(0) \rangle - \langle \mathcal{O}_{i}(t) \rangle \langle \mathcal{O}^{\dagger}_{j}(0) \rangle~,
\end{equation}
where $\mathcal{O}^{\dagger}_{j}(0)$ and $\mathcal{O}_{i}(t)$ are interpolating operators that create and annihilate a state respectively. After inserting a complete set of eigenstates of the Hamiltonian, the two-point function becomes 
\begin{equation}
C_{ij}(t) \sim \sum_{n} e^{-E_{n}t}~,
\end{equation}
where $E_{n}$ is the energy of the $n^{th}$ state and the sum is over a discrete set of states due to the finite volume.

In order to maximise the spectroscopic information we can extract from the two-point functions, we employ a variational technique \cite{Michael:1985ne, Luscher:1990ck} which amounts to solving a generalised eigenvalue problem
\begin{equation}
C_{ij}(t)v^{n}_{j} = \lambda^{n}(t,t_{0})C_{ij}(t_{0})v^{n}_{j}~, 
\end{equation}
where $C_{ij}(t)$ is given by equation \eqref{eqn:correlation_function}, and $n = 1, 2, \dots , N$, where $N$ is the number of interpolating operators used in \eqref{eqn:correlation_function}. The solution procedure yields a set of eigenvalues $\{ \lambda^{n}(t,t_{0}) \}$, ordered such that $\lambda^{1}(t,t_{0}) > \lambda^{2}(t,t_{0}) > \dots > \lambda^{N}(t,t_{0})$, and a corresponding set of eigenvectors $\{v^{n}_{j}\}$ for each $t$ and reference time-slice $t_{0}$. At large enough values of $t$, the eigenvalues are proportional to $e^{-E_{n}(t-t_{0})}$. Hence, the energy, $E_{n}$, of state $n$ can be extracted via
\begin{equation}
m^{eff}_{n} \equiv a_{t}E_{n} = \ln \left( \frac{\lambda^{n}(t,t_{0})}{\lambda^{n}(t+1,t_{0})} \right)~, 
\end{equation}
at large enough values of $t$, where $m^{eff}_{n}$ is known as the \textit{effective mass} of state $n$. 

Of course, the quality of spectroscopic extraction will depend on the type and number of interpolating operators used in a given channel. In the scalar (Higgs) channel we use two different types of operator:~$\text{tr}\{ P \}$~and a fundamentally Higgs-like operator~$\text{tr}\{ \Phi(n_{\mu}) \Phi^{\dagger}(n_{\mu}) \}$, where the Higgs field $\Phi$ ig given by
\begin{equation}
\Phi(n_{\mu}) = \left[ \frac{1}{4N_{5}}(P(n_{\mu}) - P^{\dagger}(n_{\mu}))~,~g \right]~,
\end{equation}
and $P$ is the orbifold projected Polyakov loop around the extra dimension. In the vector (Z boson) channel we also construct two types of interpolating operator. The first of these is given by
\begin{equation}
\text{tr}\{ Z_{k} \} = \frac{1}{L^{3}} \sum_{n_{1}, n_{2}, n_{3}} \text{tr}\{ g U_{k}(n_{\mu}, k) \alpha(n_{\mu} + a\hat{k}) U^{\dagger}(n_{\mu}, k) \alpha(n_{\mu}) \}~,
\end{equation}
where we are using the $SU(2)$ projected Higgs field  $\alpha(n_{\mu}) = \Phi(n_{\mu}) / \sqrt{\text{det}(\Phi(n_{\mu}))}$ as the basic building block. We construct the second vector operator in a similar fashion, except that this time we take the orbifold projected Polyakov line, $l(n_{\mu})$, as our basic building block,
\begin{equation}
\text{tr}\{ \tilde{Z}_{k} \} = \frac{1}{L^{3}} \sum_{n_{1}, n_{2}, n_{3}} \text{tr} \{ g  l^{\dagger}(n_{\mu})  U(n_{\mu}, k)|_{n_{5} = 0}  l(n_{\mu} + a\hat{k})  U^{\dagger}(n_{\mu}, k)|_{n_{5}=N_{5}} \}~.
\end{equation}

We further increase our basis of operators in each channel through the use of two smearing techniques. Firstly, we smear the gauge links $i$ times using hypercubic smearing (HYP) \cite{Hasenfratz:2001hp}. We then construct Polyakov loops from the HYP smeared links, $P_{i}(n_{\mu})$, and apply $j$ levels of an APE-like smearing to them,
\begin{equation} 
P_{ij}(n_{\mu}) = (1 - c) P_{i(j-1)}(n_{\mu}) + \frac{c}{6} \sum_{\substack{n_{0}=n'_{0}\\ |\vec{n} - \vec{n}'| = a}} U(n_{\mu}, n'_{\mu}) P_{i(j-1)}(n'_{\mu}) U^{\dagger}(n_{\mu}, n'_{\mu})~,
\end{equation}
creating Polyakov loops of various smearing levels, $P_{ij}$ (an $SU(2)$ projection step is preformed after each iteration $j$). We use the resulting Polyakov loops for various values of $i$ and $j$ in the construction of the operators described above, increasing our operator basis significantly.

\section*{Phase Diagram}

\begin{figure}[t]
\begin{center}
\includegraphics[width=0.8\textwidth]{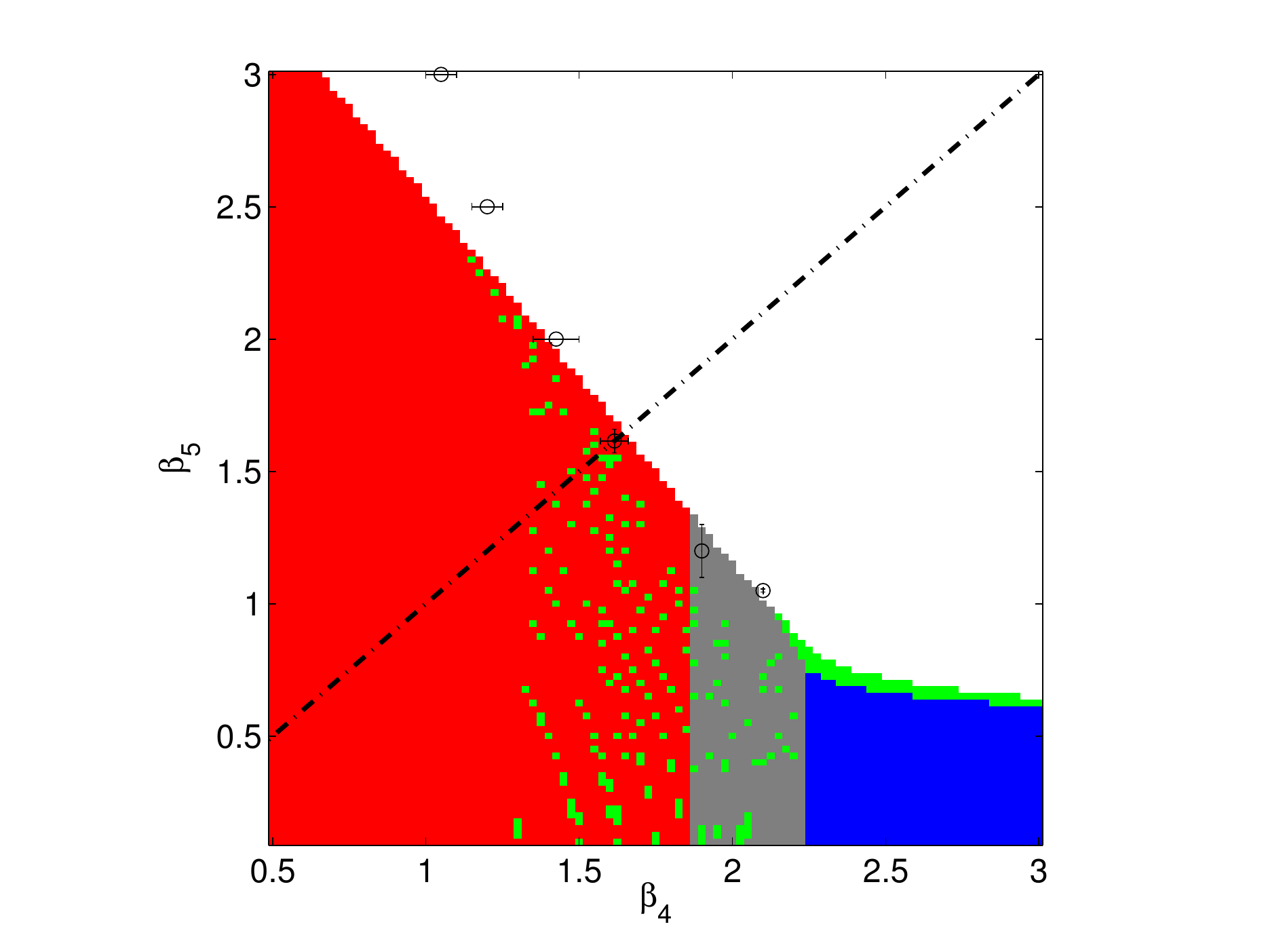}
\end{center}
\caption{\footnotesize Phase diagram of the orbifold ($N_5=4$) in the $(\beta_4,\beta_5)$ lattice gauge coupling plane. Data from the Monte Carlo phase transition (black circles) are superimposed onto the coloured phases determined from the mean-field background. The broken line represents values where the bare anisotropy $\gamma=1$. The colour coding of the phases is explained in the text.}
\label{fig:phases}
\end{figure}
This theory has three parameters; the gauge coupling $\beta$, the anisotropy, $\gamma$, which is a measure of the ratio of the lattice spacing in the four dimensions to that in the fifth dimension, and the extent of the extra dimension, $N_{5}$. In what follows we will fix the parameter $N_{5} = 4$. In Figure \ref{fig:phases} we show the phase diagram calculated from the mean-field background following \cite{Irges:2012ih}. The phase coloured red is the confined phase where the mean field background is zero everywhere. The deconfined (Higgs) phase is coloured white and it corresponds to a non-zero mean field background everywhere. The blue phase is known as the layered phase having a non-zero background within a 4-d hyperplane and a zero background along the extra dimension. The grey band represents a so-called hybrid phase, where the 4-d background is non-zero at the fixed points of the orbifold and zero elsewhere. The green points represent that the phase was not determined at the given $(\beta_4,\beta_5)$ values. Superimposed onto the mean-field phase diagram are black points which represent the locations of the bulk phase transition as determined via Monte Carlo simulations of the system. From these simulations it was found that the bulk phase transition is of first order \cite{Irges:2006hg}.

\section*{Spectrum}

\begin{figure}[t]
\begin{center}
\includegraphics[width=0.95\textwidth]{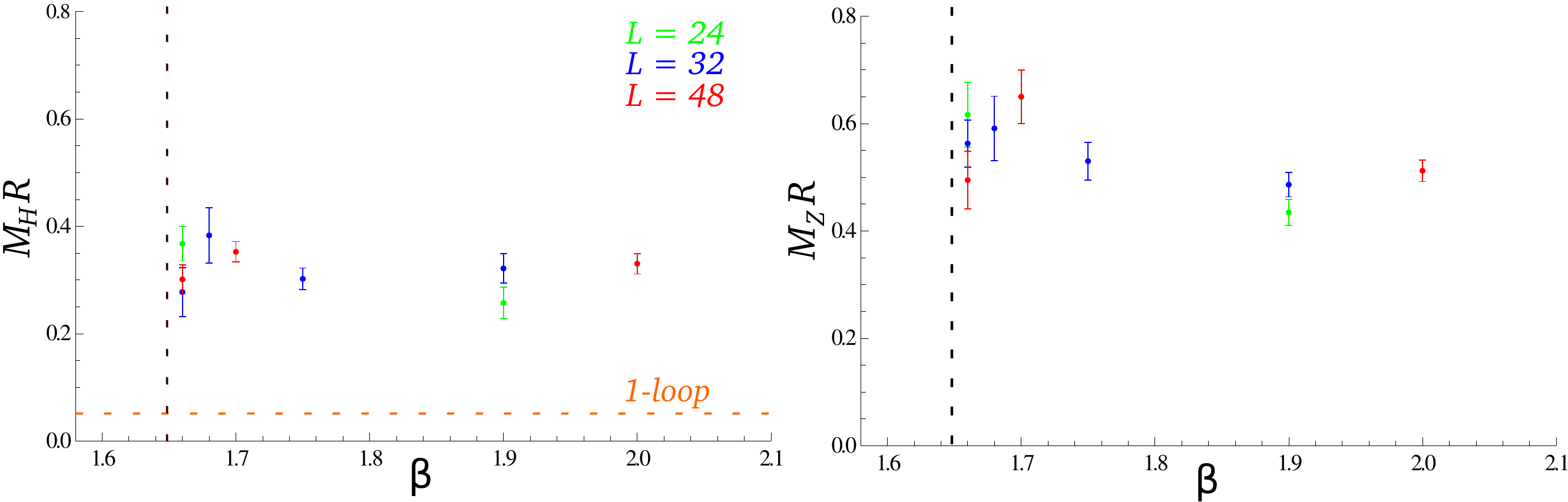}
\end{center}
\caption{The left (right) panel shows the mass of the Higgs-like (Z-like) boson along the isotropic line for $N_{5}=4$ in units of the radius of the extra dimension. The colour coding represents the extent of the spatial dimensions of the lattice.}
\label{fig:spectrum_iso}
\end{figure}
In the left (right) panel of Figure \ref{fig:spectrum_iso} we show the calculated mass of the Higgs-like (Z-like) scalar (vector), in units of the radius of the extra dimension, along the isotropic $\gamma=1$ line depicted in Figure \ref{fig:phases}. The colour coding of the points represents the extent, $L$, of the spatial dimensions of the lattice. Finite volume effects appear to be under control, at least for the $L=32, 48$ lattices. The value of the Higgs mass is significantly larger than the 1-loop calculation (broken orange line) suggesting that non-perturbative effects in this system are large. For all of the values of $\beta$ shown, the Z boson mass is larger than that of the Higgs. However, mean-field calculations find that the correct hierarchy of masses is achieved in the $\gamma < 1$ region of the phase diagram \cite{Irges:2012ih}, suggesting that the method of dimensional reduction is that of localisation (compactification is expected to occur for $\gamma > 1$).

\begin{figure}[t]
\begin{center}
\includegraphics[width=0.7\textwidth]{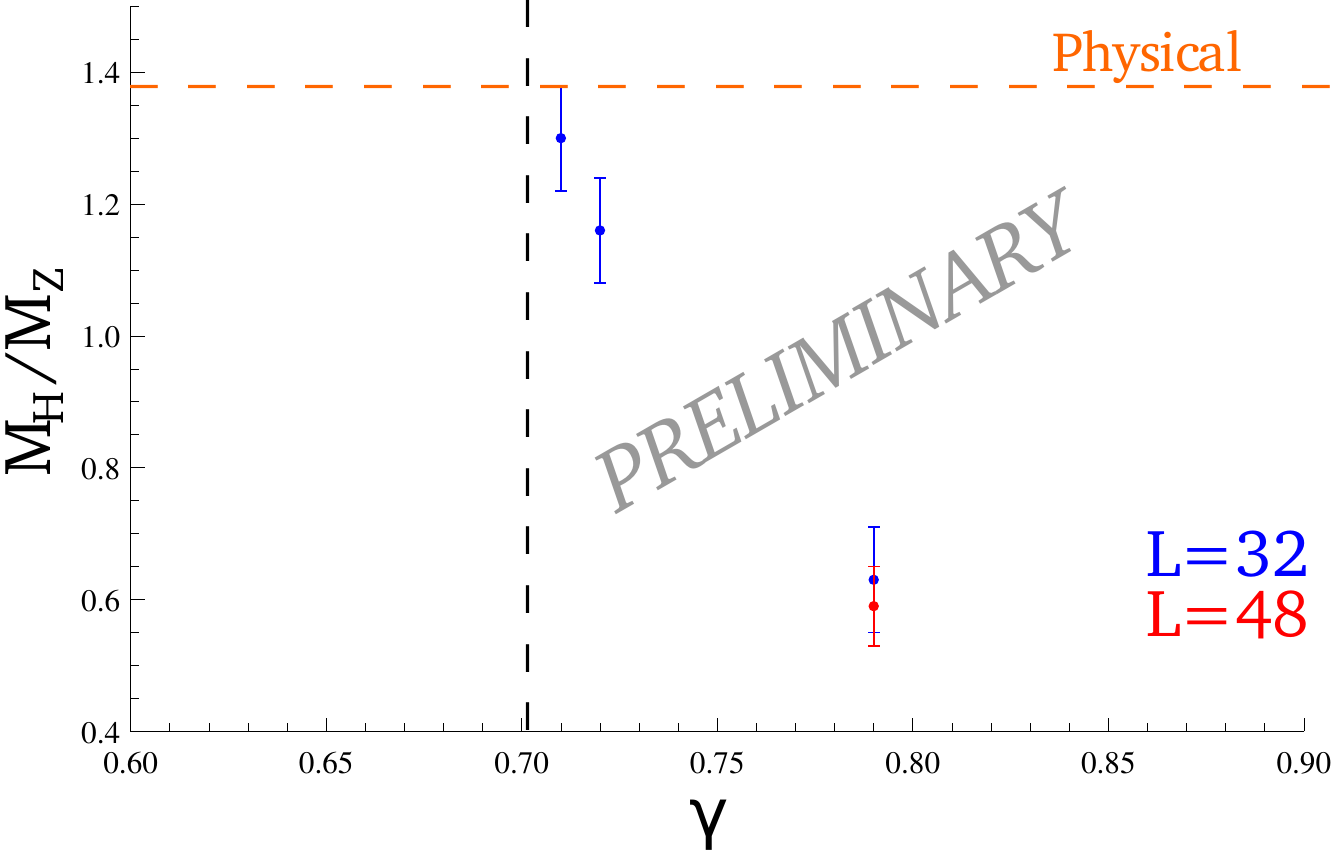}
\end{center}
\caption{The ratio of the mass of the Higgs-like particle to that of the Z-like particle when approaching the bulk phase transition keeping $\beta_4=2.1$ and lowering the value of $\beta_{5}$ for $N_{5}=4$. The colour coding indicates the extent of the spatial dimensions of the lattice and the broken black line indicates the value of the bulk phase transition.}
\label{fig:spectrum_lowgamma}
\end{figure}

Motivated by the mean-field result, we fix $\beta_4 = 2.1$ and calculate the spectrum lowering the anisotropy $\gamma$, of which Figure \ref{fig:spectrum_lowgamma} shows the result. As before, the colour coding represents the extent of the spatial dimensions of the lattice and finite volume effects appear to be under control. The broken black line indicates the value at which we find the bulk phase transition. Approaching this transition we see an increase in the ratio $\rho=M_{H}/M_{Z}$ and at $\gamma \sim 0.71$ it is $1.30(9)$, consistent with the experimental value of $\rho_{expt} =1.38$.

Further calculations in the $\gamma < 1$ regime are currently under way, along with a first exploration at $\gamma > 1$ where it is expected that the method of dimensional reduction is compactification and that the spectrum resembles that of a Kaluza-Klein theory. 

\section*{Acknowledgements}

This work was funded by the Deutsche Forschungsgemeinschaft (DFG) under contract KN 947/1-2. In particular G. M. acknowledges full support from the DFG. The Monte Carlo simulations were carried out on the Cheops, one of the supercomputers funded by the DFG at the RRZK computing centre of the University of Cologne and on the cluster Stromboli at the University of Wuppertal and we thank both Universities.

\end{document}